\documentclass[aps,showpacs]{revtex4} 
\usepackage{times}
\usepackage{graphicx}
\usepackage{amsfonts}
\usepackage{amssymb}
\usepackage{amsmath}
\usepackage{amsopn}
\usepackage{xspace}
\usepackage{bm}
\usepackage{array}
\usepackage{epsfig}
\newcommand{\be}{\begin{equation}}
\newcommand{\ee}{\end{equation}}

\newcommand{\ber}{\begin{eqnarray}}
\newcommand{\eer}{\end{eqnarray}}

\newcommand{\na}{\nabla}
\newcommand{\bra}{\langle}
\newcommand{\ket}{\rangle}
\begin{document}
\title{The fractal geometry and the mapping of Efimov states to Bloch states}

\author{ Ehoud Pazy}
\email{ehoudpazy@gmail.com}
\affiliation{Department of Physics, NRCN, P.O.B. 9001, Beer-Sheva 84190, Israel}
\date{\today}


\begin{abstract}

Efimov states are known to have a discrete real space scale invariance, working in  momentum space we identify the relevant discrete scale invariance for the scattering amplitude  defining its
 Weierstrass function as well. Through the use of the mathematical formalism for discrete scale invariance for the scattering amplitude we identify the scaling parameters from the pole structure of the 
corresponding zeta function, it's zeroth order pole is fixed by the Efimov physics. The corresponding geometrical fractal structure for Efimov physics in momentum space is identified as a  ray across a logarithmic spiral. This geometrical structure also appears in the physics of atomic collapse in the relativistic regime connecting it to Efimov physics. Transforming to logarithmic variables in momentum space we map the three-body scattering amplitude into Bloch states and the ladder of energies of the Efimov states are simply obtained in
 terms of the Bohr-Sommerfeld quantization rule. Thus through the mapping the complex problem of three-body short range interaction is transformed to that of a non-interacting
 single particle in a discrete lattice.

\end{abstract}
 
\maketitle
\section{Introduction}
\label{sec:Intro}
The quantum physics of  three body  resonantly interacting particles is known to generate a universal hierarchy of shallow three body-states states as originally established by Efimov \cite{Efimov, Efimov1} for identical bosons.
 For a fairly recent review of the more general problem, e.g. treating the three fermion state, the effect of dimensionality as well as an update on the experimental side see \cite{Naidon17}. For particles interacting through
 short-range attractive interactions that are nearly resonant the number of bound three particle state,  Efimov states, becomes infinite if at least two of the two-body interactions have an infinite s-wave scattering length. 
Aside from the amazing fact that these states are formed in a regime where two particles can not bound, exhibiting
what is referred to as Borromean binding, these three-particle bound states exhibit  a discrete scaling symmetry. The discrete symmetry is manifested in  the size $R_n$ and binding energy $E_n$ of the $n-$th Efimov
 state which scales spatially as $R_n=\lambda R_{n-1}$ and correspondingly for the energy as $E_n=\lambda^{-2}E_{n-1}$ with respect to the underlying $(n-1)$ Efimov state. For the homonuclear Efimov states the scale factor is given by
 $\lambda_0=e^{\pi/s_0}$ were $s_0=1.00624$ is a universal constant.

Having eluded experimental verification for over three decades since their initial prediction, Efimov states were experimentally observed \cite{experiment} and their universality has been demonstrated
(see Ref. \cite{expRev} for an experimental review). The experimental observation was facilitated by the  ability to greatly enhance the scattering length in atomic systems via Feshbach resonances attaining low-energy
universality in atomic few-body systems which has lead to a surge of theoretical and experimental effort in the study of few body physics. Despite the large flow of experimental papers in which evidence for Efimov states was
 presented for homonuclear systems the experimental confirmation of the discrete scaling trait was long forthcoming. In 2014 two Efimov states of Cs atoms were subsequently observed
 \cite{{Huang14}} and three Efimov states were experimentally detected  concurrently in  a heteronuclear Li-Cs mixtures at the University of Chicago \cite{Shih-Kuang14} and in Heidelberg \cite{Pires14}.

Originally of interest to the cold atom and nuclear physics communities the importance of Efimov physics has been extended to many other fields. Only lately aspects of Efimov physics 
have been shown to be of relevance in solid state systems particularly to topological semi-metals  and graphene. More specifically 
it is has been shown to be related to quasi-Rydberg resonances in graphene and to physics of atomic collapse \cite{Ovdat17} as well as to interaction of electron
with an impurity in a Dirac semi-metal \cite{Zhang18}. Recently the self similarity of Efimov states had been shown to extend to
the time domain  and ideas for observing the phenomena in cold atom systems \cite{Gao19} and trapped ion systems have been suggested \cite{Lee19}.
Amazingly also a connection between Efimov physics and the the binding of three stranded DNA, which is a classical system, was  established \cite{Maji10}, exhibiting
in a sense what has been referred to as biological Efmov effect.

On the theoretical side, Efimov first obtained his original solution employing hyper-spherical coordinates \cite{Efimov, Efimov1}. Though originally met with skepticism 
the validity of Efimov's result was established both analytically and numerically by Amado and Noble \cite{Amado71,Amado72}. Later the three body system with short 
range interactions was addressed in terms of an effective field theory (EFT) formulation of the problem \cite{Nuclphys99,Bedaque98,Bedaque99,TheoRev}. Extensions
of the theory beyond the three body system case have also been considered \cite{Castin10,Bazak17}.
The scaling behavior of the Efimov states is most evident when transforming the problem to an  effective Schr\"odinger equation for a single particle
in an inverse square potential \cite{TheoRev}. The obtained Schr\"odinger equation is invariant under a continuous scale transformation however the 
problem is ill defined at short scales since the Hamiltonian is not self adjoint. To remedy this issue one needs to impose a boundary condition, resulting in a remarkable
result the boundary condition breaks the continuous scale invariance spontaneously into a discrete scale symmetry. Such a  breaking of a continuous scale symmetry in the quantum domain is a
manifestation of non-relativistic scale anomaly \cite{Ovdat18}. Another notable issue is that the scaling factor is universal in the sense that it is independent of the chosen boundary
conditions in terms of renormalization group (RG), this effect is tied to the limit cycle behavior of the renormalization group flow equation \cite{Albeverio81}. It was originally proposed by
Wilson \cite{Wilson71} that the renormalization group equations can in addition to fixed point solutions also admit limit cycle solutions corresponding to a discrete scale invariance with respect
to a scaling factor corresponding to the oscillation period. Such solutions exhibit a log-periodic dependence as a function of the characteristic scale. However limit cycle RG solutions are
quite rare and Efimov physics is probably the most notable example of such a solution \cite{Albeverio81,TheoRev}. 

Originally considered as an oddity in the energy spectrum of three particles with short-range interactions Efimov physics  has long been established to have profound connections to a wide range of
physical problems. On the theoretical  side the  Efimov spectrum has been shown to form a geometric series corresponding to an infinite number of weakly bound states with an
 accumulation at the zero energy threshold. The associated Efimov states thus posses a discrete scale invariance which is connected to a limit cycle RG limit.
In this paper we focus on the geometrical aspect of Efimov states by employing the mathematical formalism for functions with discrete scale invariance (DSI), we identify
the relevant scaling parameters and establish the appropriate Weierstrass function. The geometrical underlying fractal structure is identified and finally we use these 
observations to greatly simplify the problem by mapping it to that of a Bloch state. 

The reminder of this paper is organized as follows.
In Sec. \ref{sec:EFT} we give a brief introduction to Efimov physics, employing the EFT formulation. Specifically the scattering amplitude is calculated via EFT.
The mathematics of  functions possessing a DSI is shortly reviewed in Sec.\ref{sec:DSI}. The connection between
scattering amplitude and DSI is established in Sec.\ref{sec:Efimov_DSI} based on a Neumann series expansion and  the corresponding Weierstrass function for the Efimov scattering amplitude 
is identified as well. In this section the underlying fractal structure of the scattering amplitude is established to be a ray across a spiral.
Transforming to logarithmic variables Efimov states are mapped to Bloch
states in Sec.\ref{sec:Bloch} and  the Efimov spectrum is obtained from
the Bohr-Sommerfeld quantization rule. A brief discussion of the real space formalism is presented in Sec.\ref{sec:frac} . 
Results are discussed and summarized in Sec.\ref{sec:sum}.

\section{Efimov physics in terms of EFT in a nutshell}
\label{sec:EFT}
It is convenient to introduce Efimov physics in terms of EFT, in this section we do so briefly for the sake of clarity. The presentation closely follows the method described in Refs. \cite{ Bedaque99,TheoRev}.
The basis for the EFT is the most general  non-relativistic Lagrangian for a boson field $\psi$  with mass $M$, which is invariant under small-velocity Lorentz transformation and  parity,
\be
\label{eq:Lagrangian}
\mathcal{L} =\psi^{\dagger} \left ( \imath \partial_0 +{\na^2 \over 2 M} \right) \psi- {C_0 \over 2}(\psi^{\dagger}\psi)^2-{D_0 \over 6}(\psi^{\dagger}\psi)^3 + \cdots,
\ee
where $C_0$ and $D_0$ are the bare low-energy coupling constants (LECs) for the two and three body interactions respectively. To make the theory renormalizable it  is defined up to an ultraviolet cut-off,
$\Lambda$. Introducing  a dummy field, $d$, corresponding to a local operator which annihilates two bosons at a point, the Lagrangian can be re-written without the three and the two body
contact interaction terms. 
\be
\label{eq:diLag}
\mathcal{L} =\psi^{\dagger} \left ( \imath \partial_0 +{\na^2 \over 2 M} \right) \psi+ \Delta (d^{\dagger}d)- {g \over \sqrt{2}}(d^{\dagger}\psi \psi +h.c. )+h (d^{\dagger}d \psi^{\dagger}\psi +h.c.)
\ee
Employing perturbation theory through a diagrammatic expansion in terms of Feynman diagrams,
the first step involves calculating the dressed propagator for the dummy field, obtained by summing bubble diagrams to all orders. The next step is obtaining an integral  for the Fourier transform
of the amputated connected part of the Green’s function  $\bra 0 |T(d \psi d^{\dagger} \psi^{\dagger}) |0 \ket$, resulting in a Skorniakov-Ter-Martirosian (STM)  equation \cite{STM} for the scattering amplitude 
\begin{equation}
\label{eq:integral}
a_{sc}(p)=K(p,k)+{2  \lambda \over \pi}\int_0^{\Lambda} dq K(p,q){ q^2 \over q^2-k^2 -i\epsilon} a_{sc}(q),
\end{equation}
where $k (p) $ is the incoming (outgoing) momentum and $\lambda=1$ for the bosonic case.
For the case when $a_2$, the two particle s-wave scattering length, satisfies the following condition,  $1/a_2 \ll p\ll \Lambda$ and $k \sim1/a_2$, the Kernel in the integral Eq. (\ref{eq:integral}) is approximated by
 
\be
\label{eq:kernel}
K (p,q)={2\over \sqrt{3} q}\ln{({q^2+pq+p^2\over q^2-pq+p^2})}.
\ee
 The main contribution to the integral comes from momenta in the  intermediate region $1/a_2 \ll q\ll \Lambda$
for which the following simplified integral equation can be obtained
\be
\label{eq:simp_int}
a_{sc}(p)={4 \over \sqrt{3} \pi} \int_0^\infty {dq \over q}a_{sc}(q)h(q,p)
\ee
where 
\be
\label{eq:transO}
h(q,p)=\ln{({q^2+pq+p^2\over q^2-pq+p^2})}.
\ee
The scale invariance of the above Eq. (\ref{eq:transO}) suggest a power-law solution $a_{sc}(p) \sim p^s$ to Eq. (\ref{eq:simp_int}). Such a power-law solution requires that $s$ satisfy the following condition
\be
\label{eq:conds}
1-{4 \over \sqrt{3} \pi} \mathcal{M}_{h(q,p)}(s)=0.
\ee
where $\mathcal{M}_{h(q,p)}$ is the Mellin transform with respect to the variable, $q$  and the transform is defined as 
\be
\label{eq:Mellin}
\mathcal{M}_{f(x)}(s)=\int_0^\infty dx \ x^{s-1} f(x).
\ee
There are two imaginary solutions to Eq. (\ref{eq:conds}) given by
\be
\label{eq:s}
s=\pm \imath s_0 
\ee
with $s_0 \simeq 1.0064$. Thus the solution to Eq. (\ref{eq:simp_int}) is given by
the  linear combination of 
\be
\label{eq:solution}
a_{sc}(p)=c_{+}p^{\imath s_0}+c_{-}p^{-\imath s_0}
\ee
where $c_{\pm}$ are constants.
In the above mentioned region, $1/a_2 \ll p \ll \Lambda$, the phase of $a_{sc}(p)$ is well determined 
\be
\label{eq:lopersol}
a_{sc}(p)=Acos( s_0\ln{p \over \Lambda}+\delta),
\ee 
where $A$ and  $\delta$ are some undetermined constants. It should be noted that the form of Eqs. (\ref{eq:solution},\ref{eq:lopersol}) already suggests the underlying one dimensional periodic physics
on which we will elaborate on in Sec. \ref{sec:Bloch}. In addition the log periodic solution (\ref{eq:lopersol}) is a known property of functions with a DSI. In the following section we will
present, for the sake of clarity, a brief sketch of the properties of functions with DSI as a basis for establishing a connection to  the three-body scattering amplitude.

\section{A brief introduction to Discrete scaling functions}
\label{sec:DSI}
To help express the scattering amplitude in terms of functions with a DSI we  briefly digress in this section 
to describe the mathematical properties of functions with a discrete symmetry.  
The material in this section is a short summary of the introductory material found in Ref. \cite{Akkermans13} combined with 
Ref. \cite{Sornette98}. In general a function with a discrete scale invariance will obey the following equation
\begin{equation}
\label{eq: discrete_sym}
f(x)=g(x)+{1 \over b}f(ax).
\end{equation}
Were $a$ and $b$ are scaling parameters and $g(x)$ an initial function. 
In describing physical systems Eq. (\ref{eq: discrete_sym}) typically becomes exact only asymptotically, however for systems defined on a on regular geometrical 
fractals such as the Cantor set, the Sierpinsky Gasket, etc., it is exact on all scales for nearest neighbor interactions
\cite{Gluzman02}. Thus functions satisfying Eq. (\ref{eq: discrete_sym}) can usually be associated with a fractal structure 

The general solution to the above Eq. (\ref{eq: discrete_sym}) is given by
\be
\label{eq:formal_sol}
f(x)=x^{\ln{b}/\ln{a}}G({\ln {x} \over\ln{a}})
\ee
where $G(x)$ is a periodic function of, $x$, with period unity. 
A formal iterative solution for Eq. (\ref{eq: discrete_sym}) is given in the form 
\begin{equation}
\label{eq:itrative_solution}
f(x)=\sum_{n=0}^{\infty}b^{-n}g(a^nx).
\end{equation}
In the specific case were $g(u)=\cos(u)$ were $u=a^nx$ one obtains the most well known example of a continuous function
which (for $ab>1$) is nowhere differentiable introduced by Weierstrass \cite{Weierstras1872} and named after him 
\be
\label{eq:Weierstrass}
W(x)=\sum_{n=0}^{\infty}({1 \over b})^{n}\cos(a^n\pi x).
\ee
In the regime were the Weierstrass has a fractal structure one can associate with it a fractal dimension 
\be
\label{eq:Dimension}
D_H=2+ {\ln b \over \ln a }
\ee 
For physical systems these sort of solutions, Eqs. (\ref{eq:itrative_solution},\ref{eq:Weierstrass}) are usually considered as an asymptotic case obtained for expansions close
enough to a fixed point
\cite{Gluzman02}.
Preforming a Mellin transform (defined in Eq. (\ref{eq:Mellin})), on Eq. (\ref{eq:itrative_solution}) one obtains the zeta function for $f(x)$ 
\be
\label{eq:zeta}
\zeta_f(s)={b a^s \zeta_g(s) \over1-ba^s}.
\ee
The zeta function has the following definition 
\be
\label{eq:zeta_def}
\zeta_f(s)={M_f(s) \over \Gamma(s) },
\ee
where $\Gamma$ is the Euler Gamma function.

The pole structure of  $\zeta_f(s)$ in Eq. (\ref{eq:zeta}) is composed of poles of the analytical function $g$ 
which generally occur for integer values and contribute only to the regular  part
of $f$ and the poles resulting from the DSI which are given by the solutions $s_n$ of 
\be
\label{eq:pole_struc}
ba^s=1 
\ee
Specifically 
\be
\label{eq:poles}
s_n=-{\ln{b} \over \ln{a}}+{2 \pi \imath n \over \ ln{a}}
\ee
The $n=0$ case gives the real power law solution $f(x)=Cx^\alpha$ to the homogeneous version of Eq.  (\ref{eq: discrete_sym}), i.e $g(x)=0$, were $\alpha=-{\ln{b} \over \ln{a}}$
which from Eq. (\ref{eq:Dimension}) can also be presented as $\alpha=2-D_H$.

\section{Describing the scattering amplitude in terms of  a DSI  function}
\label{sec:Efimov_DSI}
Armed with the mathematics for DSI functions we can now return to the physics of  the scattering amplitude, described in Sec \ref{sec:EFT}, and attempt to reinterpret it  
in terms of self similar structures. Specifically it is of interest to find what the  parameters $a$ and $b$ which define the  scaling in Eq. (\ref{eq: discrete_sym})
 are for the case of Efimov physics. Furthermore it is of interest to unmask the underlying fractal geometrical structure at the basis of the scattering amplitude. 

\subsection{Introducing a condition on the Neumann series solution to the EFT integral equation}
\label{subsec:Neumann}
In constructing a mapping between the integral equation for the scattering amplitude (\ref{eq:integral}) to Eq. (\ref{eq: discrete_sym}) which defines functions possessing  a
DSI symmetry, we start by formally considering the  iterative  solution to the integral Eq. (\ref{eq:integral})  as a Neumann series as
\be 
\label{eq:solNeu}
a_{sc}(p)=\sum_{n=0}^{\infty} \tilde{\lambda}^n \psi_n(p)
\ee
where $\tilde{\lambda} = {2 \lambda /\pi}$, $\psi_n(p)=\int_0^{\Lambda} dk \int_0^{\Lambda} \cdots \int_0^{\Lambda} \cdots K(p,q_1)K(q_1,q_2)... K(q_n,k)dq_1 \cdots dq_n, $ and $K(p,k)$ is defined by Eq. (\ref{eq:kernel}).
 Eq. (\ref{eq:solNeu}) can be rewritten in the form of the iterative solution Eq. (\ref{eq:itrative_solution}) for a DSI function  under
the following condition
\be
\label{eq:map_cond}
\int_0^{\Lambda} dq K(p,q)K(q,k)={2 \over \sqrt{3}} K(a_ep,k),
\ee
where $a_e$ is a parameter, currently unfixed, which can later be identified with the corresponding  DSI parameter in Eq. (\ref{eq: discrete_sym}). The equality in condition Eq. (\ref{eq:map_cond}) is evident when applying a Mellin transformation to
both sides (see Eq. (\ref{eq:conv})).
Essentially Eq. (\ref{eq:map_cond}) is equivalent to the homogeneous form of  Eq.(\ref{eq: discrete_sym}) in terms of the Greens function
since $\int  dq K(p,q)K(q,k)=K(p,k)$ and employing this in Eq. (\ref{eq:map_cond}) one obtains 
\be
\label{eq:sim}
K(p,k)={2 \over \sqrt{3}} K(a_e p,k).
\ee
Thus by comparing  the above Eq. (\ref{eq:sim}) to the homogeneous form of Eq. (\ref{eq: discrete_sym}) one obtains the scaling parameter $b \sim{ \sqrt{3}/2}$ (up to the
${\tilde{\lambda}}^{-1}=\pi/2$ factor).
 Alternatively under the condition in Eq. (\ref{eq:map_cond}) , Eq. (\ref{eq:solNeu}) transforms to
\be
\label{eq:selfsim}
a_{sc}(p)=\sum_{n=0}^{\infty}{\left( {2\tilde{\lambda} \over \sqrt{3}}\right) }^n \int_0^{\Lambda} dk K({a_e}^n p,k).
\ee

From  Eqs. (\ref{eq: discrete_sym},\ref{eq:itrative_solution})  one can discern the scaling parameter $b_e={ \sqrt{3} \pi/4}$ and the periodic function
\be
\label{eq:periodic}
g_e(p)=\int_0^{\Lambda} dk K( p,k),
\ee
for the Efimov scattering amplitude. The index $e$ was added to identify functions and parameters relevant to
Efimov physics.

In  Eq. (\ref{eq: discrete_sym}) the two DSI scaling parameters $a$ and $b$ were assumed to be fixed defining the pole structure of the corresponding zeta function, Eq. (\ref{eq:poles}).
For the scattering amplitude for the Efimov physics these two parameters need to be inferred. Whereas the value of $b_e$ is directly deduced from Eq. (\ref{eq:selfsim}), 
the value of the parameter, $a_e$, needs to be established from  the pole structure of the corresponding zeta function. Since the Efimov scattering amplitude obeys a DSI
symmetry its parameters obey Eq. (\ref{eq:pole_struc}) and since for Efimov physics  $s=\pm s_0$ is fixed therefore $a_e$ is automatically defined by $b_e$ and $s_0$ as 
\be
\label{eq:parameters}
a_e=\exp{(\imath \ln b_e/s_0)} ;  \  \ b_e={ \sqrt{3} \pi \over4}.
\ee
To obtain this result in a more rigorous manner, one starts out by multiplying both sides of condition (\ref {eq:map_cond}) by $p^{(s-1)}$ and integrating over $p$ thus obtaining
\be
\label{eq:conv}
{4 \over 3}\mathcal{M}_{h(p,q)}(s)\mathcal{M}_{h(q,k)}(s)= {4 \over 3}{a_e}^{-s}\mathcal{M}_{h(p,k)}(s),
\ee
where the l.h.s of the equation is obtained from the Mellin convolution theorem. The condition for the Efimov scattering amplitude stated in  Eq. (\ref{eq:conds}) 
 can also be written as
\be 
\label{eq:condr}
\mathcal{M}_{h(q,k)}(s)={1 \over b_e}
\ee
Keeping in mind that $\mathcal{M}_{h(x,y)} \equiv \mathcal{M}_{h(x/y)}$, one obtains on inserting the above condition Eq. (\ref{eq:condr}) into Eq. (\ref{eq:conv}),
exactly the condition for the pole structure of a DSI function defined in Eq. (\ref{eq:pole_struc}). However since this condition has
to hold specifically for $s=\pm \imath s_0$ the value of $a_e$ is thus fixed, but in contrast to the DSI functions
discussed in the previous Sec. \ref{sec:DSI}, the parameter $a_e$ for the scattering amplitude is complex. Furthermore one should note that 
the parameter $a_e$ is also a function of the scale parameter $b_e$ in difference to what is commonly the case as they are usually
fixed independently. These properties have consequences 
regarding the self similar geometric structure which underlays the Efimov physics, as will be discussed shortly. As a side remark
we note that the condition $b_ea_e^{\pm s_0}=1$  for the Efimov states can be  directly obtained from the self similar nature of
the scattering amplitude, $a_{sc}(a_e p)=b_e a_{sc}(p)$ by simply assuming a power law solution $a_{sc}(p) \sim p^s$.

It is important to note that for the Efimov scattering amplitude, aside from the two poles $s=\pm \imath s_0$, the pole structure can take on 
the following values
\be
\label{eq:svalues}
s_n=\pm \imath s_0+{2 n \pi s_0 \over \ln b_e}
\ee
 This pole structure should be contrasted with the DSI pole structure introduced in Eq. (\ref{eq:poles}) in which the real and imaginary terms are interchanged.

\subsection{The corresponding Weierstrass function}
\label{subsec:frac}

Given the values of the DSI parameters Eq. (\ref{eq:parameters}) and the function series for the scattering amplitude, Eq.(\ref{eq:selfsim}), we can proceed and demonstrate 
that the scattering amplitude can be expressed as a Weierstrass function Eq. (\ref{eq:Weierstrass}).  We begin by evaluating $g_e(p)$ defined in Eq. (\ref{eq:periodic}) by estimating the sum
 in Eq. (\ref{eq:selfsim}) by a saddle point approximation. Taking the derivative in terms of $n$ and defining $u={a_e}^nx$ we obtain 
\be
\label{eq:sp}
(\ln b_e) g_e(u)+(\ln a_e)u{g_e}'(u)=0.
\ee 
Inserting the value of $a_e=\exp{(\imath \ln b_e/s_0)}$ from Eq.(\ref{eq:parameters}) into the above Eq. (\ref{eq:sp}) simplifies the expression to
\be
\label{eq:sp_sim}
{g_e'(u) \over g_e(u)}=-{\imath s_0 \over u}.
\ee
From which $g_e(u)$ is obtained as
\be
\label{eq:g}
g_e(u)=\exp{(-{\imath s_0 \over u})}
\ee
however a solution with the opposite sign is also admissible, Eq. (\ref{eq:s}), thus
\be
\label{eq:gf}
g_e(u)=2\cos({s_0 \over u}).
\ee
Moreover from the definition of $ K( p,k)$ in Eq. (\ref{eq:kernel}) one obtains that
$g_e(p)=g_e({1 \over p})$ and so we obtain
\be
\label{eq:final}
g_e(u)=2\cos({u \over s_0}).
\ee
By employing the above result Eq. (\ref{eq:final}), and inserting the value of $g_e(p)$ into Eq. (\ref{eq:selfsim}), with the definition of $b_e$ from Eq. (\ref{eq:parameters}), 
the Efimov scattering amplitude is expressed by a Weierstrass function, (Eq. (\ref{eq:Weierstrass}))
\be
\label{eq:scatWeier}
a_{sc}(p)= 2 \sum_{n=0}^{\infty}b^{-n}_e \cos ({a^n_ e p \over s_0})
\ee

\subsection{The underlying fractal structure}
\label{subsec:spiral}

The DSI for the Efimov scattering amplitude is defined by the two scaling parameters Eq. (\ref{eq:parameters}). Each iterative transformation multiplies the function by
a constant factor  $ b_e={ \sqrt{3} \pi /4}$ and introduces a phase $a_e=\exp{(\imath \ln b/s_0)}$. It is natural to connect this symmetry to a one dimensional 
scattering system however such a mapping does not convey the full underlaying geometric fractal like structure at the basis of the DSI. To uncover 
the relevant underlying geometry, one should extend the one dimensional form into the plane. Since the scale parameter $a_e$  is complex it is convenient
to consider the extension in term of polar coordinates. We consider the  scaling parameter  $b_e$ as a radius vector which is a function of the angle $b_e(\theta)$.  As such this also introduces
an angle dependence for $a_e(b(\theta))$. Since the periodicity requires $a_e(\theta+2\pi)=a_e(\theta)$ one obtains the following constraint on $b_e$ such that
\be
\label{eq:constraint}
\ln b_e(\theta+2 \pi)=\ln b_e(\theta)+2 \pi s_0.
\ee
To obey the above restriction, Eq. (\ref{eq:constraint}), $b_e$ should be of a logarithmic spiral form
\be
\label{eq:be}
b_e(\theta)=b_0e^{\theta s_0}
\ee
where now $b_0=b_e$. 

An alternative method to obtain the same result is obtained by considering Eq. (\ref{eq:pole_struc}) for $s=\imath s_0 $. In this case the condition can be written as
\be 
\label{eq:cond}
{a_e}^{\imath s_0}b_e=1 .
\ee
By the following representation  $a_e(\theta)=|a_e| \exp{[\imath(\theta-\theta_0)]}$ one obtains
\be
\label{eq:cond1} 
b_e e^{\imath s_0 \ln (|a_e| e^{\imath (\theta-\theta_0)})}=b_e e^{\imath s_0 \theta \ln |a_e|-s_0( \theta-\theta_0)}.
\ee
Since $|a_e|=1$  from Eq.(\ref{eq:parameters}) one obtains again Eq. (\ref{eq:be}) by identifying $b_0=\exp{(-s_0 \theta_0)}$.

In general the logarithmic spiral  in terms of polar coordinates $(r,\theta)$ is defined by two parameters $c$ and $k$ such that
$r=c \exp{(\alpha \theta)}$. It posses the following interesting property which is highly relevant to the DSI as well as the Efimov physics, the logarithmic spiral  is a self similar structure in the sense that 
scaling by a factor $\exp({2 \pi \alpha})$ results in the same structure, as a result  any ray from each center meets the spiral at distances which are a geometric progression.
Thus one can infer  from Eq. (\ref{eq:be}) that in the case corresponding to Efimov physics the spiral parameter $\alpha$ can be identified as $\alpha= s_0$. Thus the fractal nature of the
Efimov scattering amplitude is identified as a logarithmic spiral which is obtained by rotating by $2\pi$ and stretching. Where the scaling parameter, $a_e$,  is responsible for the rotation
and the scaling parameter, $b_e$, for the stretching or shrinking of the function. The related geometric procedure can be viewed in the following way. Moving along the spiral an angle of
$2 \pi$ and rescaling the radius by, $b_e$, one returns to the initial position.

\subsection{Obtaining the expression for the scattering amplitude from the DSI formulation}
\label{subsec:scatamp} 

Based on the  knowledge gained regarding functions with DSI  (see Sec. {\ref{sec:DSI}) the expression for the Efimov scattering amplitude Eq. (\ref{eq:lopersol}), previously obtained by solving the relevant integral
STM Eq. (\ref{eq:integral}), can now be realized directly by the mapping. Starting with the general solution of Eq. (\ref{eq: discrete_sym}) given in Eq. (\ref{eq:formal_sol}) in which
 $G(y)$ is an arbitrary periodic function of its argument $y$ with period 1, considering the scaling parameter $a_e$ as given by
Eq. (\ref{eq:parameters}) with two possible complex values for  $ s=\pm \imath s_0$ and expressing the solution in terms of a logarithmic variable $ \tilde{x}=\ln p$
\begin{equation}
\label{eq:general_sol}
f_{\pm}(\tilde{x})=e^{\pm \imath s_0\tilde{x}}G({\tilde{x}\over \ln a_e})
\end{equation}
The above solution Eq.(\ref{eq:general_sol}) involves the two scales of the three-body physics, the short length scale physics is defined by the periodic 
function $G(y)$ and the long length scale physics is described by the plane wave. Taking the limit in which one can replace
the periodic function $G(y)$ by a constant, the long scale solution is given by the solution is obtained by a linear combination of the $e^{\imath s_0 \tilde{x}}$ 
and $e^{-\imath s_0 \tilde{x}}$. Re-expressing the result  in terms of $p$ again, one obtains the known result  \cite{Bedaque99} for the scattering amplitude
$f_{sc}(p)=A\cos[s_0 \ln(p/\Lambda)+\delta]$, where $\delta$ is some phase to be determined by the boundary conditions. We will  proceed below to demonstrate that
under this different view point a reinterpretation of the three-body physics can be given in terms of Bloch functions  on a one dimensional lattice. In this reformulation the  root of the zeta function, $s_0$,  is analogous to the wave 
number and $ \tilde{x}=\ln p$ is analogous to the spatial dimension along the one-dimensional lattice.

\section{Mapping Efimov states to Bloch states on a one dimensional lattice}
\label{sec:Bloch}

In this section we use the DSI formalism to map the three body Efimov physics into that of a single particle confined to a one dimensional discrete lattice. 

\subsection {Identifying the crystal momentum and lattice constant}
\label{subsec:Latt}

Eq. (\ref{eq:selfsim}) for the three body Efimov scattering amplitude, in terms of out-going momenta $p$, was identified with the iterative solution 
Eq. (\ref{eq:itrative_solution}) of the STM Eq. (\ref{eq:integral}). Through the similarity between the two equations the two scaling parameters
$a_e$ and $b_e$, where identified (see  Eq.(\ref{eq:parameters})). Based on this  mapping the Efimov scattering amplitude can be expressed in terms of
the general solution for function obeying a DSI, Eq. (\ref{eq:formal_sol}), using the scaling parameters and transforming to a logarithmic variable $\tilde{x}=\ln p$
\be 
\label{eq: Efimovg}
a_{sc}(\tilde{x})=e^{\imath k \tilde{x} }G [{  \tilde{x} \over  L }]
\ee
where we defined $k \equiv \imath{ (\ln b_e /\ln a_e)}$, $L=-\imath\ln a_e$ and $G(x)$ is a periodic function with period 1.

\begin{table}[h!]
  \begin{center}
    \caption{Efimov, DSI, Bloch correspondence}
    \label{tab:table1}
\renewcommand{\arraystretch}{2}
    \begin{tabular}{|c|c|c|c|}
     \hline
  \textbf{State}     & \textbf{DSI} & \textbf{Efimov} & \textbf{Bloch}\\ 
   \hline
   \textbf{Coordinate} &   $  x $ & $\tilde{x}=\ln p$ &  $x$\\
      \hline
      \textbf{Poles} &  $s_n=-{\ln{b} \over \ln{a}}+{2 \pi \imath n \over ln{a}}$ & $s_n=\pm \imath s_0+{2 n \pi s_0 \over \ln b_e}$ & $k={2 \pi n \over L}$\\ 
      \hline
       \textbf{Relevant Fuction} & Weierstrass function  & scattering amplitude  &  Bloch wave function\\ 
       \hline 
     &  $W(x)=\sum_{n=0}^{\infty}({1 \over b})^{n}\cos(a^n\pi x)$ & $a_{sc}(\tilde{x})=e^{\imath k \tilde{x} }G [{  \tilde{x} \over  L }]$ & $\psi(x)=e^{\imath k {x} }u(x)$\\ 
     \hline
     \textbf{Symmetry}  &   scale invariance &  scale invariance ($p$)   &  translation invariance \\
    &   &   translation invariance ($\tilde{x}$)  &  \\ 
   \hline 
 \textbf{Geometrical Structure}& Fractal& Spiral & One-D Lattice \\
    \hline 
 \textbf{Relevant Transform}& Mellin & Mellin & Fourier series \\
    \hline
    \end{tabular}
  \end{center}
\end{table}
The similarities between the Efimov scattering amplitude, functions with a DSI and Bloch functions are stated in
Tab. \ref{tab:table1}. To better understand the correspondence between the Efimov scattering amplitude to the physics of a particle
on a one dimensional lattice we consider the pole structure of a function obeying a DSI as defined in Eq. (\ref{eq:poles}) and its analog  in terms of Efimov
 physics, Eq. (\ref{eq:svalues}), expressing  these in terms of $k$ and $L$, 
\be
\label{eq:crystalmom}
s=\imath k_0+{2 \pi n \over L},
\ee
were the following identifications were considered 
\ber
\label{eq:correspondence}
 s_0 & \leftrightarrow & k_0  \nonumber \\
 s_0/\ln b_e =-\imath/\ln a_e& \leftrightarrow & 1/ L.
\eer
In this form the connection of $s_n=(2 \pi n s_0 /\ln b_e)$ to a crystal momentum is self-evident. 
It should be noted that whereas the pole structure for the DSI functions as well as for the Efimov scattering amplitude
was obtained via a Mellin transform, the quasi-momentum for the Bloch states is obtained by a Fourier series expansion. However 
the two transforms can be connected if one considers the analytical continuation for the  Mellin transform for a logarithmic variable. 

The connection between the relevant underlaying symmetries, the DSI invariance of the Efimov physics to the translation symmetry of 
the one dimensional lattice system becomes clear by considering Eq. (\ref{eq:correspondence}). One should note that considering
a logarithmic variable  $\tilde{x}=\ln p$ multiplication of the momentum $p$ in the Efimov scattering amplitude 
by a constant $a_e$, translates in terms of the logarithm variable, $\tilde{x}$,  to a  shift of $\tilde{x}$ in Eq. (\ref{eq: Efimovg}) by $\ln a_e$. 
This shift results only in a phase $\exp{(\imath  k L)}$ since the function $G(x)$ in Eq. (\ref{eq: Efimovg}) is a periodic function with period 1. The phase,
according to the identifications in Eq. (\ref{eq:correspondence}),  is $s_0L=\ln b_e$. Making it clear that requiring that the Efimov scattering
amplitude as described in Eq. (\ref{eq: Efimovg}) be viewed as Bloch function on a lattice with a lattice period, $L=\ln b_e/ s_0$
defined by the condition $\psi(x+L)=\exp{(\imath k L)} \psi(x) $ is equivalent to requiring DSI symmetry.  
Thus identifying the scattering amplitude Eq. (\ref{eq: Efimovg}) with 
a Bloch function through the definitions in Eq. (\ref{eq:correspondence}) leads directly to the connection between the two underlaying symmetries, i.e.,
multiplying $p$ by $a_e$ is equivalent to dividing or multiplying the function by $b_e$ which exactly defines the DSI. 

Having  established  the analogy to Bloch states
one can now view the Efimov physics in terms of logarithmic variable  $\tilde{x}$ as describing the physics of particle on a one-dimensional lattice, where 
$\ln a_e$, plays the role of the lattice constant $L$ ,and $s_0$ plays the role of the particles momentum. The power-law ansatz
 $a_{sc}(p)\sim p^s$ suggested as a solution to Eq. (\ref{eq:simp_int}) \cite{Bedaque99} simply maps into a plane wave
solution $a_{sc}(\tilde{x}) \sim \exp{(\imath \tilde{x}s_0)}$ where $s$ is limited to the values $\pm \imath s_0$.

To make this analogy more explicit we consider the scattering amplitude, Eq. (\ref{eq:lopersol}), expressed in terms of the  logarithmic variable $\tilde{x}=\ln p$
we can now reinterpret it in terms of a Bloch state 
\be
\label{eq:trans}
a_{sc}(p)=Acos( s_0\ln{p \over \Lambda}+\delta) \rightarrow a_{sc}(\tilde{x})=Acos( k_0\tilde{x}+\delta'),
\ee
where $\delta'$ is a phase.
In this representation where the incoming momenta (in the Efimov description) plays the role of the spatial coordinate and $s_0$ plays the role of an effective
crystal momentum. 

\subsection {Obtaining the Efimov spectrum from the Bohr-Sommerfeld quantization}
\label{subsec:Bohr-Sommerfeld}
Employing the connection between  Efimov physics to Bloch functions and having identified the corresponding momentum through Eq. (\ref{eq:correspondence}), the Efimov spectrum 
can now simply be  obtained from the Bohr-Sommerfeld quantization rule 

\begin{equation}
\label{eq:Bohr_Sommerfeld}
s_0 \int_{\tilde{x}_1}^{\tilde{x}}d\tilde{x}=(n+\delta)\pi,
\end{equation}
where $\delta$ is a phase resulting from the boundary conditions. Keeping in mind that the spatial coordinate is actually the logarithm of the momentum in the Efimov
description $\tilde{x}=\ln p$, expressing this above integral Eq. (\ref{eq:Bohr_Sommerfeld})  in terms of $p$, performing the integration and employing the viral theorem we obtain the Efimov spectrum
\begin{equation}
\label{eq:Efimov_spectrum}
E_n=-{p_{*}^2\over m}e^{-2 \pi n /s_0},
\end{equation}
where $p_{*}=e^{-\delta \pi/ s_0}\ln{\tilde{x}_1}$.

\section{Real space formalism}
\label{sec:frac}

In this section we analyze Efimov physics in real space thus we are able to connect it with the physics of the relativistic atomic collapse which is defined through the same fractal geometry, that of a ray across a logarithmic spiral.

\begin{figure}
\centering
\includegraphics[scale=0.4]{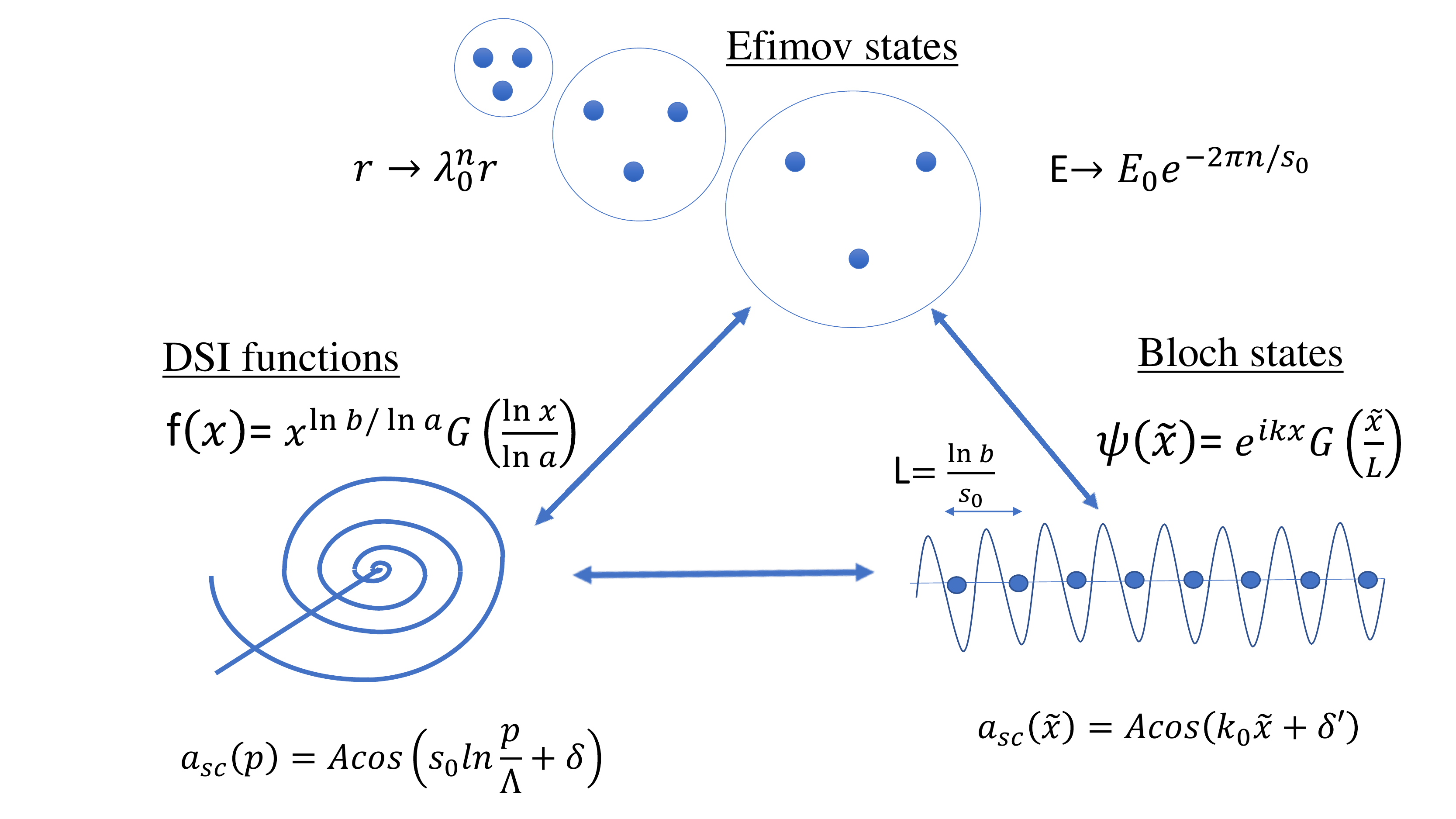}
\caption{ Starting with the wave function for the Efimov states in real space transforming to momenta space and considering the scattering amplitude the logarithmic spiral self similar geometry is
obtained. Transforming to exponential coordinates maps the physics
into the physics of Bloch states on a one-dimensional lattice.}
\label{fig:1}
\end{figure}

\subsection{The Efimov connection to DSI in real space}
\label{subsec:real}
In considering an operator with a Weierstrass spectrum $\gamma^n$ where $n$ is an integer it was shown \cite{Berry80} that 
the Schr\"odinger equation  for the inverse square potential of the form 
\be
\label{eq:potential}
U(x)= -{A\over  x^2}
\ee
 has an infinitely deep spectrum
with states clustering at $E=0$
\be
\label{eq:Wspec}
E_n=-E_0 \gamma^n
\ee
referred to as a Weierstrass spectrum. When one requires that two solutions with energies  $E_1, E_2$ are
orthogonal  the ratio between these energies is given by
\be
\label{eq:ratio}
{E_1 \over E_2}=\exp\left ({2 \pi n \over \sqrt{A-{1 \over 4}}}\right ),
\ee
where $n$ is an integer.
To obtain the Weierstrass spectrum, Eq. (\ref{eq:Wspec})  the constant $A$, in Eq. (\ref{eq:potential}), defining the potential takes the value
\be
\label{eq:Wcond}
A={1 \over 4}+{4 {\pi}^2 \over {\ln}^2{\gamma}}.
\ee
The above result relates directly to Efimov physics since in the real space calculations using hyper-spherical coordinates for the three 
body problem Efimov physics emerges from a Schr\"{o}dinger equation with an 
inverse squared potential \cite{TheoRev}. Given for convenience in coordinates in which $\hbar=1$
\be
\label{eq:schrodinger}
-{1 \over 2 \mu}\left [  \left ( {d^2 \over dR^2}  \right)+{{s_0}^2+1/4 \over R^2} \right ]\psi(R)=E \psi(R)
\ee
where $R\equiv={2 \over 3}(r_{12}^2+r_{13}^2+r_{23}^2)$, and $r_{ij}$ are the relative particle coordinates $\mu$ is the effective mass and
$E$ is the energy of the bound state. Comparing Eqs. ({\ref{eq:potential},\ref{eq:Wcond}) to the above Eq. (\ref{eq:schrodinger}) one obtains the following  identification 
\be
\label{eq:identication}
s_0 \leftrightarrow \pm {2 \pi \over \ln \gamma}.
\ee

Considering a semi-classical type solution using units in which the mass of the particles is $m=1$ \cite{Nielsen99} the action given by
\be 
\label{eq:action}
S=\pm \imath \int^{R_0} dR' k(R') 
\ee
where $k(R)=\sqrt{[E- V(R)]}$ , and $V(R)=(s_0^2+1/4)/R^2$ which can be identified with $U(x)$ in Eq. (\ref{eq:potential}) by employing the connection in
 Eq. (\ref{eq:identication})  and applying it to Eq. (\ref{eq:Wcond}). In the region of small $R$ where  $E$ can be neglected the boundary conditions for the integral
 are given by the turning point $R_0$ which are  the roots of $s_0^2+1/4=0$. 
\be 
\label{eq:realspacesol}
\psi(R) \approx A_{+}e^{({1 \over 2}+\imath s_0)\ln R}+A_{-}e^{({1 \over 2}- \imath s_0)\ln R}
\ee
where $A_{+}$ and $A_{-}$ are constants.
 
According to  the semi-classical  Bohr-Sommerfeld approximation for the phase difference between the two cases should be 
quantized such that
\be
\label{eq:BohrSommerfeldquant}
2 s_0 \ln (R)= 2\pi n
\ee
The Bohr-Sommerfeld condition, Eq. (\ref{eq:BohrSommerfeldquant}), thus defines the discrete periodicity of $k(R) \sim \ln R$ which is at the heart of Efimov physics and DSI functions.
The above condition Eq. (\ref{eq:BohrSommerfeldquant}) is equivalent  to Eq. (\ref{eq:Bohr_Sommerfeld}) and hence also
leads to the spectrum in Eq. (\ref{eq:Efimov_spectrum}).

Similar physics is described in Ref.  \cite{Levitov07} in which the problem of atomic collapse in the relativistic regime is analyzed. It is worth mentioning since in
 this case the spiral geometry is clearly evident. The collapsing trajectories for an electron into the nucleus  are best defined through the quasi-classical  radial momentum equation.
\be
\label{eq:atomic_collapse}
p_R^2=v_F^{-2} \left (E+{Ze^2 \over R} \right )^2-{M^2 \over R^2}
\ee
where $R$ is the radial distance to the nucleus $p_R$ is the radial momentum, $e$ the electron charge, $Z$ is the atomic number, $M$ the electron angular momentum and $v_F$ the Fermi velocity.
The spiral trajectories are directly manifest in the atom collapse for the case $M<Ze^2/c$ for $E>0$ and the ray across the spiral
is manifested for the tunneling case in which  $M<Ze^2/c$ for $E<0$. The resulting spectrum for the problem is equally spaced on a log scale
\be
\label{eq:ACspectrum}
E_n \approx {Ze^2\over r_0} e^{-\pi \hbar n /\gamma}
\ee
where $r_0$ is a lattice cut-off, $\gamma=(M_c^2-M^2)^{1\over 2} $ and $M_c=Ze^2/v_F$ is the critical angular momentum separating the falling trajectories from the stable trajectories. 

It should also be noted that the Efimov physics is a special case in a more general set of problems which involve a singular inverse square  potential. This set of equations also has  a very similar dependence on a  critical scale as the relativistic atomic collapse case, such that in order to obtain a discrete energy spectrum
it is needed that in the potential, Eq.(\ref{eq:potential}), $A>A_{cr}$ where $A_{cr}={(d-2)}^2/4$  and $d$ is the dimension of the problem. Stated this way the discrete spectrum is given by
\be 
\label{eq:discretespec}
E_n \propto e^{2 \pi n \over \Lambda}
\ee
where $\Lambda=\sqrt{A-A_{c}}$ and the results where presented for units such that $\hbar=1$.
\section{Summary and Discussion}
\label{sec:sum}

Efimov physics has been shown to be of importance to a wide set of physical problems ranging from ultracold atomic gases \cite{Kraemer06} through nuclear physics \cite{Kunitsk15} to recently condensed matter systems \cite{Wang18}
as well as to biological systems \cite{Maji10}. On the theoretical side it has even been considered for more than three particles for which it was originally formulated \cite{Castin10,Bazak17}.
In this work we have examined the
less studied aspects of the geometric and more specifically the fractal properties of the Efimov scattering amplitude. Whereas initially  it was known from the start that the Efimov spectrum  
forms a geometric series corresponding to an infinite number of weakly bound states which posses a discrete scale invariance, later also connected to a limit cycle RG limit, here we
have demonstrated how one can apply the mathematical formalism for functions with DSI to the Efimov scattering amplitude. 
There are however some differences, in contrast  to the theory described for functions possessing DSI symmetry in which one considers the two scaling parameters $a$ and $b$ as independent and fixed
from the geometry determining the value of $s_0$ the zeroth pole of the corresponding zeta function, in the case of Efimov physics $s_0$ is imaginary and is obtained as a solution for a transcendental equation  and $b_e$ is fixed and real. Solutions can
then be found by considering complex values for $a_e$. Using the DSI mathematical formalism we identified
the relevant scaling parameters and established the corresponding Weierstrass function.
 
A fractal is an iterative structure. Famous examples are the triadic Cantor set obtained iteratively by dividing  a segment into there parts removing the middle third part and continuing the 
process with the remaining segments. Another example is the Sierpinski gasket in which  an equilateral triangle is subdivided recursively into smaller equilateral triangles. In regards to 
the Efimov scattering amplitude  we have shown that the relevant fractal structure is a logarithmic spiral which is obtained by rotating by $2\pi$ and stretching thus leaving the
shape invariant. From the physics stand point Efimov physics is an example of a scale anomaly where the classical system posses a continuous symmetry which is broken at the quantum level to a discrete symmetry.
In terms of our reformulation of the problem a new connection between the physical symmetry of the system to the underlaying geometrical fractal can be established.
The underlying continuous  structure corresponding to Efimov physics is apparent from the fact that $a_e$ is a complex number thus multiplying by $a_e$ induces a
rotation followed by a stretching/contraction through multiplying the function by $b_e$. The one-dimensional mapping presented in the previous Sec. {\ref{sec:Bloch}} 
results by considering the discrete values along a ray through a logarithmic spiral such that the angular coordinate $\theta$ of the spiral  is  restricted to a fixed angle.
The self similarity of the scattering amplitude for  Efimov physics expressed  geometrically through the logarithmic spiral turns into translation invariance up to a phase when transforming into a logarithmic
variable allowing the mapping of the complex Efimov  physics to that of a particle on a one-dimensional lattice. The mapping allowed us to obtain the Efimov spectrum
from the  Bohr-Sommerfeld quantization rule. Returning to the field theoretic treatment one can now understand
the mapping to the one dimensional lattice already through the integral Eq. (\ref{eq:integral}) which is obtained by
summing over all terms in the perturbation theory. The fact that all term contribute "equally" (up to a phase) can be viewed as a scattering by a one dimensional lattice in which every scattering
 events introduces a phase. 

We believe the connections presented in this work offer many possible venues for extension. Specifically we speculate that through the mapping of Efimov physics to that of a one dimensional lattice a
possible connection between the  Efimov three body parameter to the geometrical Zak phase \cite{Zak} for electrons on a one dimensional lattice could be established. From a mathematical point of view the
underlying spiral identified as the underlying fractal structure for the Efimov scattering amplitude might have subtle connections to the Poincar\'e equation \cite{Derfel12}.
\\

 E.P wishes to thank Gerald Dunne, Eric Akkermans and Alxander Teplyaev for enlightening discussions on the mathematics of DSI and Vasili  Kharchenko and Jia Wang, 
for very valuable discussions of Efimov physics and especially  thank Luke Rogers for the spiral mapping idea. E.P. acknowledges the kind hospitality of the Physics Department of the University of Connecticut
in which this work originated.


\begin{thebibliography}{99}
\bibitem{Efimov}
V. Efimov, Phys. Lett. B {\bf 33}, 563 (1970).
\bibitem{Efimov1}
V. Efimov,  Yad. Fiz. {\bf 12}, 1080 (1970) [Sov. J. Nucl. Phys, {\bf 12}, 589 (1971).
\bibitem{Naidon17}
P. Naidon and S. Endo, Rep. Prog. Phys. {\bf 80}, 056001 (2017).
\bibitem{experiment}
T. Kraemer, M. Mark, P. Waldburger, J. G. Danzl, C. Chin, B. Engeser, A. D. Lange, K. Pilch, A. Jaakkola, H.-C. N\"{a}gerl and R. Grimm, Nature {\bf440}, 315 (2006).
\bibitem{expRev}
F. Ferlaino, A. Zenesini, M. Berninger, B. Huang, H.-C. N\"{a}gerl and R. Grimm, Few -Body Syst.  {\bf 51}, 113 (2011).
\bibitem{Huang14}
B. Huang, L. A. Sidorenkov, R. Grimm, and
J. M. Hutson, Phys. Rev. Lett., {\bf 112}, 190401,  (2014).
\bibitem{Shih-Kuang14}
S.-K. Tung, K. Jiménez-Garcí, J. Johansen, C. V. Parker, and C. Chin,  Phys. Rev. Lett. {\bf 113}, 240402 (2014).
\bibitem{Pires14}
R. Pires, J. Ulmanis, S. H\"{a}fner, M. Repp,
A. Arias, E. D. Kuhnle, and M. Weidem\"{u}ller, Phys. Rev. Lett.
{\bf 112}, 250404,  (2014).
\bibitem{Ovdat17}
O. Ovdat, J. Mao, Yuhang Jiang, E. Y. Andrei, E. Akkermans, Nature Communications {\bf 8},  507 (2017).
\bibitem{Zhang18}
P. Zhang, H. Zhai, Front. Phys. {\bf 13}, 137204 (2018).
\bibitem{Gao19}
C. Gao, H. Zhai, Z. Y. Shi, Phys. Rev. Lett. {\bf 122}, 230402 (2019).
\bibitem{Lee19}
D. Lee, J. Watkins, D. Frame,G. Given, R. He, N. Li, B.-N. Lu, A. Sarkar,
Phys. Rev. A {\bf 100}, 011403(R) (2019).
\bibitem{Maji10}
J. Maji, S. M. Bhattacharjee, F. Seno, and
A. Trovato,  New Journal of Physics, {\bf12}, 083057, (2010).
\bibitem{Amado71}
R.D. Amado and J.V. Noble, Phys. Lett. {\bf 35B}, 25 (1971).
\bibitem{Amado72}
R. D. Amado and J. V. Noble,  Phys. Rev. D, {\bf 5}, 1992, (1972).
E. Braaten and H. W. Hammer, Phys . Rep. {\bf 428}, 259 (2006); Ann. Phys. (N.Y) {\bf 322}, 120 (2007).
\bibitem{Bedaque98}
P.F. Bedaque, H.-W. Hammer, and U. van Kolck, Phys. Rev. C {\bf 58}, R641 (1998).
\bibitem{Bedaque99}
P. F. Bedaque , H.-W. Hammer and U. van Kolck, Phys. Rev. Lett. {\bf 82}, 463 (1999).
\bibitem{Nuclphys99}
P.F. Bedaque, H.-W. Hammer, and U. van Kolck, Nucl. Phys. A {\bf 646}, 444  (1999).
\bibitem{TheoRev}
E. Braaten and H. W. Hammer, Phys . Rep. {\bf 428}, 259 (2006); Ann. Phys. ( N.Y) {\bf 322}, 120 (2007).
\bibitem{Castin10}
Y. Castin , C. Mora,L.  Pricoupenko, Phys. Rev. Lett. {\bf105}, 223201 (2010).
\bibitem{Bazak17}
B. Bazak ,D. S. Petrov ,Phys. Rev. Lett.{\bf 118}, 083002 (2017).
\bibitem{Ovdat18}
D. K. Brattan, O. Ovdat and E. Akkermans, Phys. Rev. D {\bf 97}, 061701(R) (2018).
\bibitem{Albeverio81}
S. Albeverio, R. Høegh-Krohn, and T. T. Wu, Physics Letters A {\bf 83}, 105 (1981).
\bibitem{Wilson71}
K. Wilson, Phys. Rev. D {\bf 3}, 1818 (1971).
\bibitem{STM}
G. V. Skorniakov and K.A Ter-Martirosian, Sov. Phys. JETP {\bf 4}, 648 (1957).
\bibitem{Akkermans13}
E. Akkermans, {\it Statistical mechanics and quantum fields on fractals, Fractal geometry and dynamical
systems in pure and applied mathematics. II. Fractals in applied mathematics}, Contemp. Math., {\bf 601},
Amer. Math. Soc., Providence, RI, 2013, pp. 1–21. MR 3203824.
\bibitem{Sornette98}
D. Sornette,  Phys. Rep. {\bf 297}, 239, (1998).
\bibitem{Gluzman02}
S. Gluzman and D. Sornette, Phys. Rev. E, {\bf 65}, 036142 (2002).
\bibitem{Weierstras1872}
K. Weierstrass, " \"Uber continuirliche Functionen eines reelles Arguments, ¨
die f\"ur keinen Werth des letzteren einen Bestimmten Differentialquotienten besitzen",  K\"onigl. Akademie der Wissenschaften, Berlin, July 18,
1872. Reprinted in K. Weierstrass, Mathematische Werke II, pp. 71–74,
Johnson, New York, 1967.
1977.
\bibitem{Berry80}
M. V. Berry and Z. V. Lewis, Proc. R. Soc. A {\bf 370}, 459 (1980).
\bibitem{Nielsen99}
E. Nielsen ans J. H. Macek , Phys. Rev. Lett. {\bf 83}, 1566 (1999).
\bibitem{Levitov07}
A. V. Shytov, M. I. Katsnelson and L. S. Levitov, Phys. Rev. Lett. {\bf 99}, 246802 (2007).
\bibitem{Kraemer06}
T. Kraemer, M.  Mark , P. Waldburger, et al. Nature {\bf 440}, 315 (2006).
\bibitem{Kunitsk15}
M. Kunitski, S. Zeller, J. Voigtsberger, et al. Science {\bf 348}, 551 (2015).
\bibitem{Wang18}
H. Wang,H. Liu, Y. Li, et al., Sci. Adv., {\bf 4}, 5096 (2018).
\bibitem{Zak}
J. Zak, Phys. Rev. Lett. {\bf 62}, 2747 (1989).
\bibitem{Derfel12}
G. Derfel, P. J. Grabner, F. Vogel,
e-print: arXiv:1206.1211 (2012).

\end{thebibliography}
\end{document}